\documentclass[aps,prl,preprintnumbers,amsmath,amssymb,latexsym,nofootinbib,array,enumerate,letter,twocolumn,superscriptaddress]{revtex4-1}
\usepackage{subfigure}
\usepackage{cancel}
\usepackage{here}
\usepackage{comment,braket}
\usepackage{epsf}
\usepackage{amsmath}
\usepackage{graphics}
\usepackage{amsfonts}
\usepackage{amssymb}
\usepackage{latexsym}
\usepackage{color}
\usepackage{ulem}
\input{colordvi.tex}
\usepackage{feynmp}
\usepackage{caption}

\usepackage[pdftex]{graphicx}
\usepackage{bm}
\usepackage[pdftex]{hyperref}
\usepackage[svgnames]{xcolor}
\definecolor{phthaloblue}{rgb}{0.0, 0.06, 0.54}
\hypersetup{
    colorlinks=true,
    linkcolor=phthaloblue,
    citecolor=blue,
    filecolor=blue,
    urlcolor=phthaloblue,
    }

\newcommand{\beq}{\begin{eqnarray}} 
\newcommand{\eeq}{\end{eqnarray}}

\def\({\left(}
\def\){\right)}
\def\[{\left[}
\def\]{\right]}

\newcommand{\bel}[1] {\begin{equation}\label{#1}}
\newcommand{\beal}[1] {\begin{eqnarray}\label{#1}}
\newcommand{\be}{\begin{equation}}
\newcommand{\ee}{\end{equation}}
\newcommand{\bea}{\begin{array}} 
\newcommand{\eea}{\end{array}}

\begin{document}

\title{QCD axion dark matter and the cosmic dipole problem}

\author{Chengcheng Han}
\email{hanchch@mail.sysu.edu.cn}
\affiliation{School of Physics, Sun Yat-Sen University, Guangzhou 510275, China}
\affiliation{Asia Pacific Center for Theoretical Physics, Pohang 37673, Korea}

\date{\today}

\begin{abstract}
There is increasing evidence suggesting a discrepancy between the cosmic dipole observed in the number count of distant galaxies and the one derived from the cosmic microwave background (CMB). In this study, we investigate the possibility that the cosmic dipole problem can be addressed by considering the QCD axion, a hypothetical particle that arises from the spontaneous breaking of the Peccei-Quinn symmetry and is postulated to constitute the dark matter in our universe.
\end{abstract}

\maketitle

\section{Introduction}
Dark matter remains one of the enigmas of our universe, constituting the dominant component of observed matter and playing a pivotal role in the formation of large-scale structures. Despite its fundamental importance, the true nature of dark matter remains elusive, except for its gravitational interactions with ordinary matter. Numerous experiments have been dedicated to unraveling the particle nature of dark matter; however, thus far, no compelling evidence has emerged. Given that astronomical observations have been instrumental in providing evidence for the existence of dark matter, it is reasonable to expect that further insights from astronomy may shed light on its properties.

Interestingly, recent findings have revealed a significant discrepancy between the cosmic dipole measured from galaxy number-counts and the dipole derived from the cosmic microwave background (CMB) \cite{Wright:2010qw}. The CMB dipole arises from the motion of the solar system relative to the CMB background and indicates a velocity of the solar system of approximately 370 km/s relative to the CMB \cite{Planck:2018nkj}. Consequently, this velocity should induce a dipole in the measurements of galaxy number counts.

Comparing the cosmic dipole obtained from distant galaxy number counts with the dipole inferred from the CMB has long been employed as a means to test the cosmological principle \cite{10.1093/mnras/206.2.377, Singal:2011dy, Singal:2019pqq, Tiwari:2015tba, Rubart:2013tx, Bengaly:2017slg, Siewert:2020krp, Gibelyou:2012ri}. Studies conducted by Ellis and Baldwin \cite{10.1093/mnras/206.2.377} have demonstrated that for radio sources with identical flux density spectra described by $S \propto \nu^{-\alpha}$ (where $\nu$ denotes the frequency and $\alpha$ represents the spectral index) and an integral source count above a flux density threshold $S_*$ characterized by $dN(> S_*)/d\Omega \propto S_*^{-x}$, a dipole in the galaxy number counts arises as follows:
\begin{equation}
\frac{\Delta N}{N} = \mathbf{d_k} \cdot \hat{n}~,
\label{numbercount}
\end{equation}
where $\mathbf{d_k} = [2+x(1+\alpha)]\mathbf{\beta}$, and $\mathbf{\beta} = \mathbf{v}/v_c$ represents the relative velocity of the solar system with respect to the ``matter rest frame". By employing the velocity $\mathbf{v}$ inferred from CMB measurements in Eq.~(\ref{numbercount}) to estimate the dipole in galaxy number counts, one can compare it with the dipole obtained from actual measurements based on galaxies, thereby assessing the consistency of the cosmological principle. It is worth noting that for a comprehensive overview of previous dipole measurements, references such as \cite{Peebles:2022akh, Aluri:2022hzs} offer valuable resources.

The analysis of the CatWISE catalogue, which examines over a million quasars, has recently unveiled a discrepancy between the amplitude of the number-count dipole and the dipole derived from the CMB. This deviation has been reported to be approximately 4.9$\sigma$ in references \cite{Secrest:2020has, Secrest:2022uvx}. Such findings raise doubts regarding the validity of the cosmological principle, a fundamental assumption stating that the universe should exhibit homogeneity and isotropy on sufficiently large scales. Consequently, alternative principles have been contemplated as potential replacements for this assumption \cite{Krishnan:2022qbv}. However, before entirely discarding this fundamental principle, it is crucial to investigate whether the cosmic dipole problem can be addressed within the framework of cosmic perturbation theory. Interestingly, recent discoveries indicate that the inconsistency between the two dipoles could be alleviated if a significant dark matter isocurvature perturbation exists at super-horizon scales \cite{Domenech:2022mvt}.

In the case of isocurvature perturbations originating from the density fluctuations of dark matter, they cannot be attributed to WIMP (Weakly Interacting Massive Particle) dark matter. This is due to the fact that WIMPs would have thermally equilibrated with photons in the early universe, leading to the eradication of any initial isocurvature perturbations \cite{Weinberg:2004kf, Weinberg:2003sw}. However, axion dark matter presents a viable candidate in this context. Axions are hypothetical particles arising from the spontaneous breaking of the Peccei-Quinn symmetry, proposed as a solution to the strong CP problem \cite{Peccei:1977hh, Peccei:1977ur}. It has been long known that axion dark matter can generate isocurvature perturbations with an amplitude on the order of $\frac{H}{\pi f_a \theta_0}$, where $H$ represents the Hubble parameter during inflation, $f_a$ is the axion decay constant, and $\theta_0$ denotes the initial misalignment angle of the axion field \cite{Weinberg:1977ma, Wilczek:1977pj, Dine:1982ah, Abbott:1982af, Preskill:1982cy, Kim:2008hd, Wantz:2009it, Marsh:2015xka, DiLuzio:2020wdo, Sakharov:1994id, Sakharov:1996xg, Khlopov:1999tm}. However, it is important to note that the non-observation of isocurvature perturbations in the CMB places stringent constraints on the ratio of the Hubble parameter and the axion decay constant, yielding $H/f_a \lesssim 10^{-5}$ for $\theta_0\sim \mathcal{O}(1)$, as reported by the Planck data~\cite{Planck:2018jri}. Therefore, if axions constitute the dark matter and are responsible for the isocurvature perturbations, their properties must satisfy this constraint.

On the other hand, to address the cosmic dipole problem, a significant isocurvature perturbation is required, which contradicts the constraints imposed by CMB observations. However, if we can generate a large isocurvature perturbation at super-horizon scales that becomes sufficiently small at the scale of recombination $k_{\rm dec}$, it may be possible to reconcile this contradiction. In the following, we will present an axion model that satisfies this requirement.

This paper is organized as follows. In Section~\ref{dipole}, we will provide a brief overview of the cosmic dipole problem and present the necessary conditions to address it. In Section~\ref{axion}, we will introduce an axion model that has the potential to explain the cosmic dipole problem. The numerical calculations will be presented in Section~\ref{numerical}. Finally, in Section~\ref{con}, we will draw our conclusions based on the findings.

\section{The cosmic dipole problem}
\label{dipole}
Firstly, let us present some statistics on the calculation of the power spectrum of the cosmic microwave background (CMB). The temperature anisotropies in the CMB can be decomposed into spherical harmonics as follows:
\begin{equation}
\frac{\Delta T}{T}(\hat{\mathbf{n}}) = \sum_{l, m} a_{lm} Y_{lm}(\hat{\mathbf{n}})~,
\end{equation}
where $a_{lm}$ are the spherical harmonic coefficients given by the integral:
\begin{equation}
a_{lm} = \int d\Omega \frac{\Delta T}{T}(\hat{\mathbf{n}}) Y_{lm}(\hat{\mathbf{n}})~.
\end{equation}
The two-point correlation function for the temperature fluctuations is expressed as:
\begin{equation}
\left\langle \frac{\Delta T}{T}(\hat{\mathbf{n}}) \frac{\Delta T}{T}(\hat{\mathbf{n'}}) \right\rangle = \frac{1}{4\pi} \sum_{l=1}^\infty (2l+1) C_l P_l(\hat{\mathbf{n}} \cdot \hat{\mathbf{n'}})~,
\end{equation}
where we ignore the monopole term ($l=0$). In a homogeneous and isotropic universe, we have,
\begin{equation}
\left\langle a_{lm}^* a_{lm'} \right\rangle = \delta_{ll'} \delta_{mm'} C_l~.
\end{equation}
Typically, the dipole corresponds to $l=1$ and the quadrupole to $l=2$. The measured CMB dipole can be written as:
\begin{equation}
\frac{\Delta T}{T}(\hat{\mathbf{n}}) = d^{\text{CMB}} \hat{\mathbf{n}} \cdot \hat{\mathbf{n}}_0~,
\end{equation}
where $d^{\text{CMB}} = (1.23357 \pm 0.00036) \times 10^{-3}$ is the magnitude of the dipole and $\hat{\mathbf{n}}_0$ specifies its direction at $l=264^\circ$ and $b=48^\circ$. It is commonly assumed that this dipole arises from the Doppler effect due to the motion of the solar system with respect to the rest frame of the CMB. The relative velocity is given by:
\begin{equation}
v_o = d^{\text{CMB}} \times v_c = 369.82 \pm 0.11 , \text{km/s}~,
\end{equation}
where $v_c$ is the speed of light. For the CMB dipole, we have the relation:
\begin{equation}
\left\langle \frac{\Delta T}{T}(\hat{\mathbf{n}}) \frac{\Delta T}{T}(\hat{\mathbf{n'}}) \right\rangle = \frac{3}{4\pi} C_1 \hat{\mathbf{n}} \cdot \hat{\mathbf{n'}}.
\end{equation}
Using
\begin{equation}
\left\langle (\hat{\mathbf{n}} \cdot \hat{\mathbf{n}}_0) (\hat{\mathbf{n'}} \cdot \hat{\mathbf{n}}_0) \right\rangle = \frac{1}{3} \hat{\mathbf{n}} \cdot \hat{\mathbf{n'}}~,
\end{equation}
we get
\begin{equation}
d^{\text{CMB}} = \sqrt{\frac{9C_1}{4\pi}}~.
\end{equation}
These statistics provide the necessary background for understanding the CMB dipole and its relation to the power spectrum.

In recent studies, the measurement of kinematic dipoles from distant radio galaxies or quasars has revealed that these dipoles are 2-3 times larger than the expected value inferred from the velocity $v_o$ obtained from the CMB dipole~\cite{Secrest:2020has, Secrest:2022uvx}. This result suggests the existence of a preferred velocity given by:
\begin{equation}
v_o' = 797 \pm 87 , \text{km/s}~,
\label{velocity2}
\end{equation}
which points in a similar direction. Notably, this velocity deviates from the velocity $v_o$ inferred from the CMB dipole by approximately $4.9\sigma$. At first glance, one might be inclined to conclude that the cosmological principle should be discarded. However, before abandoning this fundamental assumption, it is crucial to investigate whether this inconsistency can be explained within the framework of the perturbed Friedmann-Lemaître-Robertson-Walker(FLRW) model. Several studies have demonstrated that if a large isocurvature perturbation exists at super-horizon scales, a portion of the CMB dipole could be intrinsic~\cite{Turner:1991dn, Langlois:1995ca, Erickcek:2008jp}. Therefore, a possible solution to reconcile the dipole discrepancy is to assume that our velocity is $v_o'$, and an intrinsic dipole cancels out a part of the kinematic dipole from the CMB, i.e.,
\begin{equation}
d^{\text{CMB}} = d_{\text{kin}}^{\text{CMB}} + D_1^{\text{CMB}} = 1.23357 \times 10^{-3}~,
\end{equation}
where $d_{\text{kin}}^{\text{CMB}} = v_o'/c$ is the kinematic dipole and the $z$-direction is chosen as the dipole direction. $D_1^{\text{CMB}}$ represents the intrinsic dipole of the CMB arising from isocurvature perturbations in the opposite direction. We can approximate it as:
\begin{equation}
D_1^{\text{CMB}} \approx -1.4 \times 10^{-3} - ({v_o' - 797  \text{km/s}})/{v_c}~.
\end{equation}
To account for the discrepancy within $2\sigma$, it would require $|D_1^{\text{CMB}}| > 8 \times 10^{-4}$. This scenario of cancellation involving an intrinsic dipole provides a potential explanation for the observed discrepancy between the CMB and kinematic dipoles.

However, before drawing this conclusion, it is necessary to investigate whether super-horizon isocurvature perturbations also impact the galaxy number-count dipole. In a recent study by~\cite{Domenech:2022mvt}, the authors examined this issue and found that isocurvature perturbations at the super-horizon scale primarily affect the CMB dipole and have only a minor influence on the galaxy number-count dipole. Consequently, the authors interpret the cosmic dipole problem as originating from a single-mode isocurvature perturbation at the super-horizon scale.

However, it is important to note that in realistic models, the power spectrum of perturbations typically exhibits a continuum rather than being dominated by a single mode. In this work, we propose an alternative explanation by attributing the cosmic dipole problem to the continuum isocurvature perturbations associated with axion dark matter. It is worth mentioning that while some articles attempt to employ axion strings to explain the CMB hemispherical anomaly, this is a distinct problem from the cosmic dipole problem we are addressing~\cite{Yang:2016wlz, Yang:2017wzh}.

According to the contributions of the isocurvature perturbation, neglecting the baryonic matter contribution, we have ~\cite{Domenech:2022mvt, Erickcek:2008jp}:
\begin{eqnarray}
\left . \frac{\Delta T}{T} \right |_{\mathbf{k}} = -\frac{1}{3} S^{ini}_{\mathbf{k}}~,
\end{eqnarray}
where $S^{ini}_{\mathbf{k}}$ represents the initial isocurvature perturbation long before recombination. This isocurvature perturbation also sources the adiabatic perturbation:
\begin{eqnarray}
\Phi_{\mathbf{k}}= -\frac{1}{5} S^{ini}_{\mathbf{k}}~.
\end{eqnarray}
It is well-known that the adiabatic perturbation generates temperature fluctuations through the Sachs-Wolfe effect:
\begin{eqnarray}
\left . \frac{\Delta T}{T} \right |^{\text{adia}}_{\mathbf{k}}= \frac{1}{3} \Phi_{\mathbf{k}}.
\end{eqnarray}
Summing these contributions, we obtain,
\begin{eqnarray}
\left . \frac{\Delta T}{T} \right |_{\mathbf{k}} = -\frac{2}{5} S^{ini}_{\mathbf{k}}~.
\end{eqnarray}
However, the super-horizon adiabatic perturbation also induces a velocity of us, which contributes to the dipole. Remarkably, its effect precisely cancels the dipole contribution from the adiabatic perturbation~\cite{Turner:1991dn, Langlois:1995ca, Erickcek:2008jp}. This explains why only the super-horizon adiabatic perturbation alone cannot generate a CMB dipole.

The power spectrum of the isocurvature perturbation, denoted as $\mathcal{P}_S(k)$, can be defined as follows:
\begin{eqnarray}
\langle S_{\mathbf{k}} S_{\mathbf{k}'} \rangle = 2\pi^2 \frac{\mathcal{P}_S(k)}{k^3} \delta(\mathbf{k-k'}) \Theta(k-k_{\text{min}})~,
\end{eqnarray}
where $k_{\text{min}}$ is the minimum comoving wave number at which the isocurvature perturbation becomes relevant.
Using this power spectrum, we can express the angular power spectrum $C_l$ as,
\begin{eqnarray}
C_l= g_l \int_{k_{\text{min}}}^\infty \frac{dk}{k}\mathcal{P}_S(k) j_l^2(k r{\text{dec}})~,
\end{eqnarray}
where $g_l=\frac{4\pi}{9}$ for $l=1$ and $g_l=\frac{16\pi}{25}$ for $l > 1$. Here, $k$ is the comoving wave number and $r_{\text{dec}} \approx 14.1$ Gpc is the comoving distance to the photon last scattering surface. To explain the cosmic dipole problem, we require
\begin{eqnarray}
\sqrt{\frac{9C_1}{4\pi}} \gtrsim 8 \times 10^{-4} \Rightarrow C_1 \gtrsim 9 \times 10^{-7}~.
\end{eqnarray}

Constraints from multipole observations, particularly the quadrupole, provide additional constraints on the power spectrum of the isocurvature perturbation. According to the Planck 2018 results~\cite{Planck:2018nkj}, the temperature variance for the quadrupole is measured to be $(200^{+ 540}_{-120}) \mu K^2$. Taking the 3$\sigma$ upper limit, we have,
\begin{eqnarray}
\frac{2(2+1)C_2}{2\pi} \lesssim 2.5 \times 10^{-10} \Rightarrow C_2 \lesssim 2.6 \times 10^{-10}.
\end{eqnarray}
Thus we have
\begin{eqnarray}
\frac{C_2}{C_1} \lesssim 3 \times 10^{-4}~.
\end{eqnarray}
It's important to note that this condition is derived from the quadrupole limit and the constraints from other multipoles are much weaker.
 
 Before presenting a specific model, let's assume that the power spectrum of the isocurvature perturbation, $\mathcal{P}_S(k)$, follows a power law given by $\mathcal{P}_S(k) = A (k/k{\rm min})^{n-1}$, where $n<1$. For $n > -1$, we can approximate the expression for the multipole moments $C_l$ as:
\begin{eqnarray}
C_l \approx g_l A (k_{\rm min} r_{\rm dec})^{1-n} c(n,l)~,
\end{eqnarray}
where $g_l$ and $c(n,l)$ are defined as before. However, this expression is only valid when $n-2+2l > -1$. If $n-2+2l < -1$, the main contribution to the integration comes from the region around $k_{\rm min}$. By using the properties of $j_l(x) \sim \frac{x^l}{(2l+1)!!}$ for $x \ll 1$, it can be shown that:
\begin{eqnarray}
C_l \approx \frac{g_l A}{|n-1+2l| ((2l+1)!!)^2} (k_{\rm min} r_{\rm dec})^{2l}~.
\end{eqnarray}
This expression is valid when $n-2+2l < -1$.

Taking the example where $n = -2$ and $k_{\rm min} r_{\rm dec} = 0.01$, we find that
\begin{eqnarray}
\frac{C_2}{C_1} \approx 1.3 \times 10^{-3}~.
\end{eqnarray}
This value is too large to satisfy the limit on the quadrupole. To address this, we can either decrease $k_{\rm min} r_{\rm dec}$ or make $n$ smaller. For instance, if we take $k_{\rm min} r_{\rm dec} = 0.0023$, we obtain
\begin{eqnarray}
\frac{C_2}{C_1} = 3 \times 10^{-4}~.
\end{eqnarray}
Assuming $A = 1$, we have $C_1 = 8.4 \times 10^{-7}$, which marginally explains the cosmic dipole discrepancy.

\section{Large isocurvature from the axion }
\label{axion}
In this section, we will demonstrate that a significant isocurvature perturbation can be generated by the axion dark matter. The energy density of the axion can be approximated as
\begin{eqnarray}
\rho_a = \frac{1}{2} m_a^2 f^2 \theta_0^2~,
\end{eqnarray}
where $f$ is the decay constant of the axion and $\theta_0$ is the initial displacement angle. In the case of the QCD axion, $f$ can typically range from $10^{11}$ GeV to $10^{15}$ GeV for a $\theta_0$ varying from $\mathcal{O}(10^{-2})$ to $\mathcal{O}(1)$. The axion mass can be estimated using the following formula,
\begin{eqnarray}
m_a = 5.7 \left(\frac{10^9~\rm GeV}{f}\right) \rm meV~.
\end{eqnarray}
For an axion-like particle (ALP), the relation between the mass and $f$ can be more relaxed.

The isocurvature perturbation of the axion can be defined as
\begin{eqnarray}
S \equiv \frac{\delta \rho_a}{\rho_a} - \frac{3}{4} \frac{\delta \rho_r}{\rho_r} \simeq \frac{\delta \rho_a}{\rho_a} = \frac{2 \delta \theta}{\theta_0},
\end{eqnarray}
where $\delta \rho_a$ and $\delta \rho_r$ are the perturbations in the axion and radiation energy densities, respectively. The amplitude of the axion perturbation $\delta \theta$ can be related to the Hubble parameter $H$ and the axion field value at a given comoving scale as
\begin{eqnarray}
\mathcal{P}^{1/2}_{\delta \theta} = \frac{H}{2\pi \varphi(k)}~.
\end{eqnarray}
Using this relation, we can express the amplitude of the isocurvature perturbation as
\begin{eqnarray}
\mathcal{P}^{1/2}_S = \frac{H}{\pi \varphi(k) \theta_0}~.
\end{eqnarray}
It is important to note that if the perturbation $\delta \theta$ is comparable to the background value $\theta_0$, we should use the effective angle $\theta_{\rm eff} = \sqrt{\delta \theta^2 + \theta_0^2}$ to calculate the power spectrum. As a result, we always have $\mathcal{P}^{1/2}_S \lesssim 1$.  From the equation above, it is evident that a large isocurvature requires $\varphi \sim H$ at the super horizon scale, while $\varphi$ should be sufficiently large at $k_{\rm dec}$ to avoid the isocurvature limit imposed by CMB observations. If the value of $\varphi$ rapidly increases to a very large value during the early stage of inflation, the tension with the isocurvature limit from CMB observations can be alleviated, as illustrated in Figure 1. In the following, we will present a model that realizes this mechanism.

The model consists of a complex scalar field $\Phi$ with a non-vanishing Peccei-Quinn charge and a potential
\begin{eqnarray}
V(\Phi)=\lambda (\Phi \Phi^\dagger -f^2/2)^2 ~.
\label{potential}
\end{eqnarray}
We define the axion field as the angular mode of $\Phi$,
\begin{eqnarray}
\Phi =\frac{1}{\sqrt{2}} \varphi \exp(i\frac{a}{f})~.
\end{eqnarray}
Assuming that initially $\varphi$ is located close to the origin but with a displacement around $H$.  Ignoring the motion of angular direction, the equation of motion for $\varphi$ is then given by,
\begin{eqnarray}
\ddot{\varphi} +3 H \dot{\varphi} + V^\prime(\varphi) =0~.
\label{EOM}
\end{eqnarray}

 \begin{center}
 \includegraphics[width=0.35\textwidth]{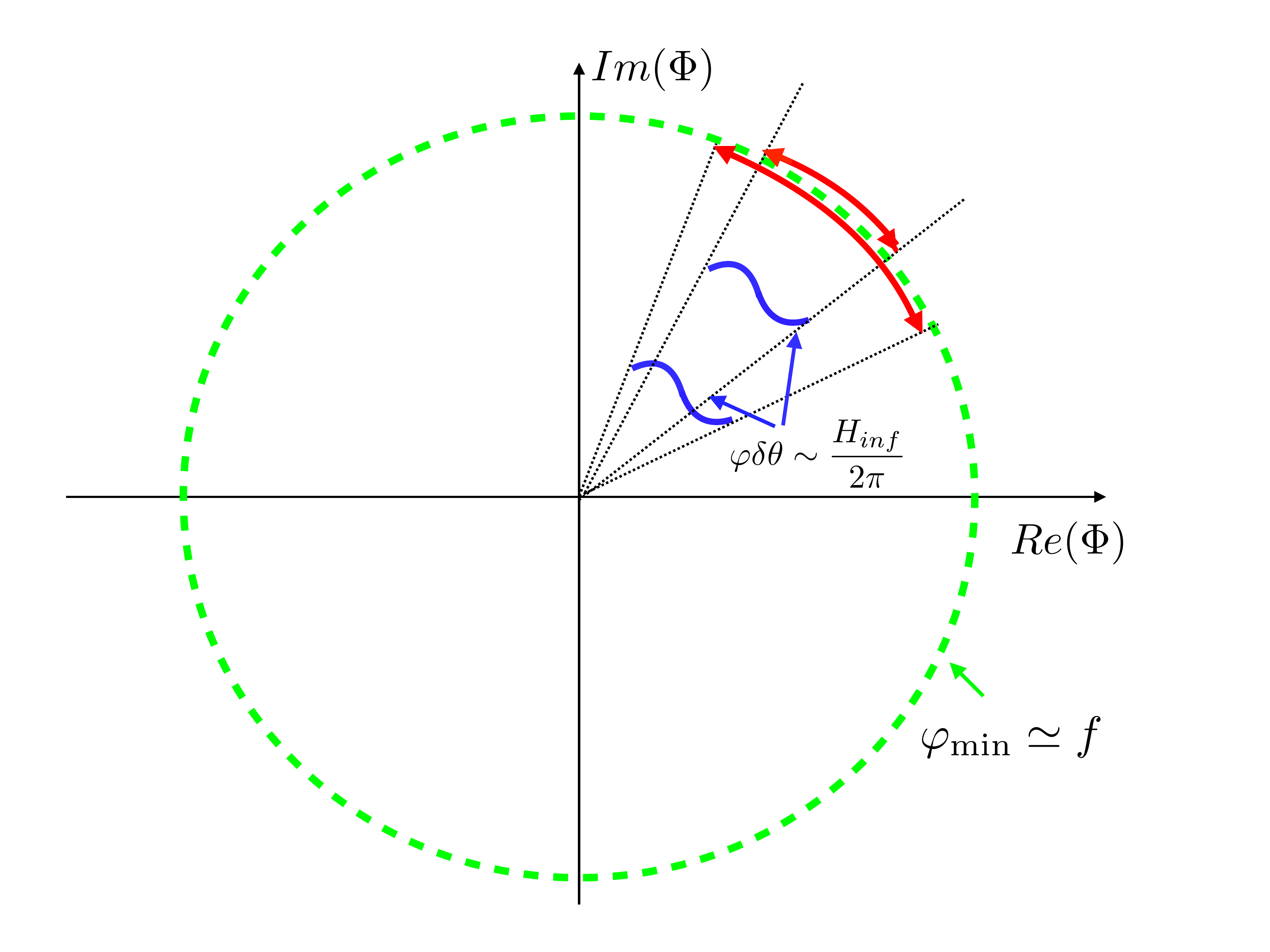}
 \captionof{figure}{Illustration for a varying isocurvature perturbation during the early universe. $\varphi$ evolves from a small value to a large value during inflation.}
\label{Dipole}
 \end{center}

The assumption that the field $\Phi$ is initially located around the origin of the potential is motivated by the presence of a coupling between the inflaton field $I$ and $\Phi$. The coupling term, $V_I = g {\Phi^\dagger \Phi I^2}$, provides an effective mass for the $\Phi$ field during inflation due to the large value of the inflaton $I$. For a positive coupling constant $g$, this effective mass is larger than the Hubble parameter, leading to a minimum of the potential at $\Phi=0$ and the preservation of the Peccei-Quinn symmetry.

During inflation, as the inflaton $I$ rolls from a large value to a smaller value, the effective mass term of $\Phi$ from $V_I$ becomes sub-dominant compared to the negative mass term from the potential in Eq.~(\ref{potential}). At a certain stage, when $I$ becomes sufficiently small, the negative mass from the potential dominates, causing the potential to develop a non-vanishing vacuum expectation value. Since the potential is flat at the top, the field $\Phi$ initially undergoes random walking and is then displaced from the origin with an amplitude around $H$.  Subsequently, $\Phi$ evolves according to the equation of motion described by Eq.~(\ref{EOM}), eventually reaching the value of $f$ after a few e-folds of inflation. If this occurs at the super horizon scale $k_{\rm min}$, the isocurvature perturbation is large at the super horizon scale and small at $k_{\rm dec}$.

\section{Numerical calculation}
\label{numerical}

For the numerical calculation,  since
\begin{eqnarray}
\mathcal P^{1/2}_S= \frac{H}{\pi \varphi(k) \theta_0}~,
\end{eqnarray}
thus
\begin{eqnarray}
\mathcal P_S(k) \approx \mathcal P_S(k_{\rm min}) \left( \frac{\varphi(k)}{\varphi(k_{\rm min})} \right )^{-2}~,
\end{eqnarray}
where $\mathcal P_S(k_{\rm min}) \simeq 1$. The evolution of $ \varphi(k)$ follows the equation of motion of~Eq.~(\ref{EOM}). Given that $k= k_{\rm min} \exp(Ht)$, we have
\begin{eqnarray}
C_l= g_l \int_{k_{\rm min}}^\infty \frac{dk}{k}\mathcal P_S(k) j_l^2(k r_{\rm dec})~.
\end{eqnarray}
In practically we can define $\tau = H t$, now above equation becomes,
\begin{eqnarray}
C_l= g_l \int_{0}^\infty {d \tau}\mathcal P_S(k_{\rm min} e^\tau) j_l^2(k_{\rm min} e^\tau r_{\rm dec})~.
\end{eqnarray}

 \begin{center}
 \includegraphics[width=0.35\textwidth]{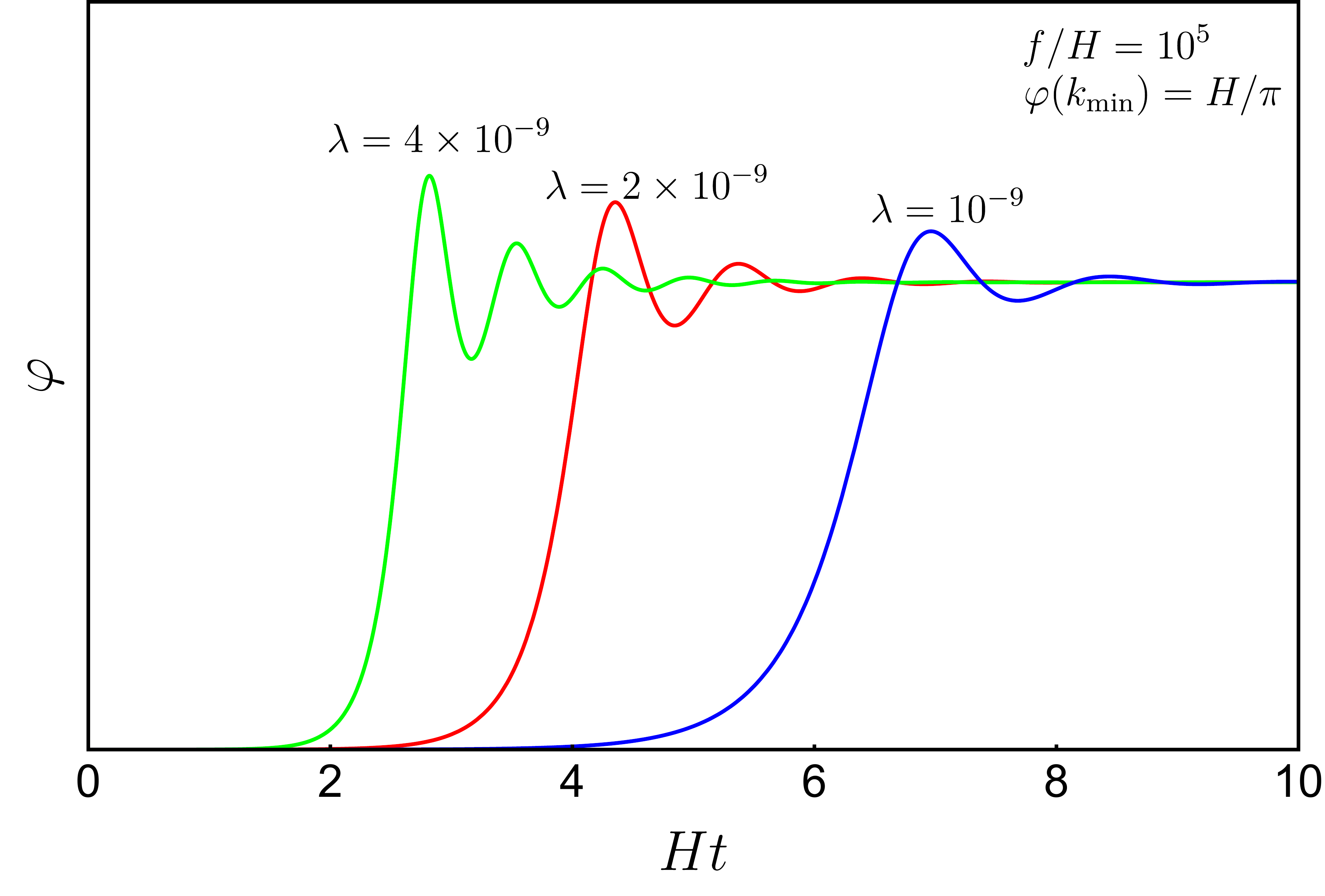}
 \captionof{figure}{The motion for the radial mode during inflation. The blue, red and green  curves correspond to $\lambda=10^{-9}, 2\times10^{-9}, 4\times10^{-9}$.}
\label{value}
 \end{center}

In Figure \ref{value}, we present the evolution of $\varphi$ for different values of the parameter $\lambda$. The blue, red, and green curves correspond to $\lambda=10^{-9}$, $2\times10^{-9}$, and $4\times10^{-9}$, respectively. The initial conditions are chosen as $\dot{\varphi}=0$, $\varphi(k_{\rm min})=H/\pi$, and $f=10^5 H$. As observed in the figure, for smaller values of $\lambda$, the field $\varphi$ remains at a small value for a longer duration. However, after a few e-folds, it eventually reaches the value of $f$. It is important to note that if $\lambda$ is too small, the field may remain at the small value for an extended period, resulting in a large isocurvature perturbation at recombination. Conversely, if $\lambda$ is too large and combined with a small $k_{\rm min}$, the field quickly reaches the value of $f$, and it becomes challenging to explain the cosmic dipole problem.

In order to address the cosmic dipole problem, we require $C_1 > 9\times10^{-7}$. However, the Planck data imposes a limit on the dark matter isocurvature perturbation, $\mathcal P_S \lesssim 0.8\times10^{-10}$ at $k_{\rm low}=0.002~{\rm Mpc}^{-1}$. To evade this limit, it is necessary to have $\varphi(k_{\rm low}) \gtrsim f/3$ for $f=10^5 H$.

In Figure \ref{parameter}, we illustrate the parameter space required to explain the cosmic dipole anomaly in the $1/(k_{\rm min} r_{\rm dec})$ versus $\lambda$ plane. The two solid curves represent $C_1=9\times 10^{-7}$ and $C_1=5.6\times 10^{-6}$, which correspond to the lower and upper limits of the intrinsic CMB dipole, respectively. The gray region is excluded due to the presence of a too-large quadrupole in the CMB, while the red region is excluded due to a significant isocurvature contribution at $k_{\rm low}=0.002~{\rm Mpc}^{-1}$.

 \begin{center}
 \includegraphics[width=0.35\textwidth]{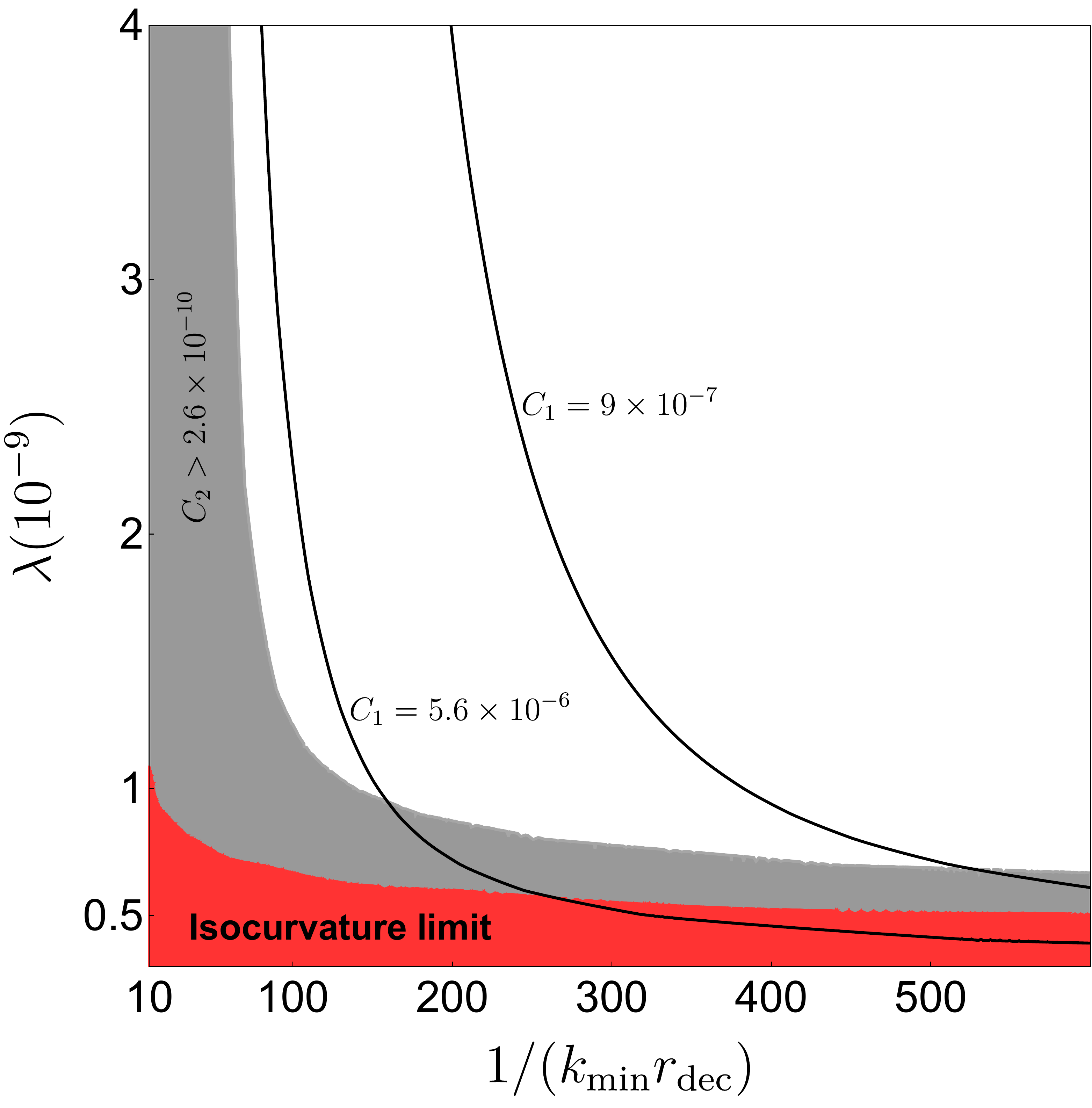}
 \captionof{figure}{Parameter space to explain the comic dipole problem. The two real lines correspond $C_1= 9\times 10^{-7}$ and $C_1= 5.6\times 10^{-6}$. The gray region is excluded because of the too large quadrupole at CMB. The red region is excluded due to too large isocurvature.   }
\label{parameter}
 \end{center}

From the plot, we observe that for $\lambda<7\times 10^{-10}$, regardless of the value of $1/(k_{\rm min} r_{\rm dec})$, the predicted quadrupole in the CMB would be too large. On the other hand, for $\lambda>4\times 10^{-9}$, values of $1/(k_{\rm min} r_{\rm dec})<250$ are necessary to generate a sufficient contribution to the cosmic dipole. It is worth noting that as $\lambda$ increases, the curve $C_1=5.6\times 10^{-6}$ eventually reaches the gray region ($C_2\lesssim 2.6\times 10^{-10}$), which sets an upper limit on $\lambda$. Numerically, we find this occurs at $\lambda=6.6\times 10^{-5}$ and $1/(k_{\rm min} r_{\rm dec})=14.6$. As a benchmark point in the model to explain the dipole problem, we consider $\lambda=2\times 10^{-9}$, $1/(k_{\rm min} r_{\rm dec})=200$, and $\mathcal P_S(k_{\rm min})\approx 1$. This yields $C_1=1.6\times 10^{-6}$ and $C_2=6.2\times 10^{-12}$.

Including the baryon matter contribution to the matter component introduces an additional factor $\left(\frac{\Omega_{cd}}{\Omega_{cd}+\Omega_b}\right)^2$ in the calculation of $C_l$. While the inclusion of this factor can affect the precise numerical values of the predicted CMB multipoles, the essential result and the overall conclusion of our analysis should remain unchanged.

\section{Conclusion}
\label{con}

In this work, we propose an axion model as a potential solution to the cosmic dipole problem. Our model predicts a large isocurvature perturbation at the super horizon scale while maintaining a small perturbation at recombination to comply with the isocurvature limit derived from CMB observations. This is achieved by allowing the radial mode of the axion field to evolve from a small value to a significantly larger value during the inflationary period. Through our analysis, we find that by choosing appropriate parameters such as $f=10^5 H$ and $7 \times 10^{-10} < \lambda < 6.6 \times 10^{-5}$, along with a suitable value of $k_{\rm min}$, it is possible to explain the cosmic dipole problem within the framework of our axion model.

It is worth noting that the cosmic dipole problem, if successfully addressed by our model, could serve as the first evidence supporting the existence of axion dark matter. This provides additional motivation for further exploration and investigation of axion physics in the context of cosmology and particle physics. Overall, our proposed axion model offers a promising avenue to resolve the cosmic dipole problem and potentially sheds light on the nature of dark matter through the presence of axions.

{\bf Acknowledgment.}---%
CH acknowledges Tsutomu T. Yanagida,  Misao Sasaki,  Kazunori Kohri and Shi Pi for helpful discussions.  CH acknowledges support from the Sun Yat-Sen University Science Foundation and the Fundamental Research Funds for the Central Universities, Sun Yat-sen University under Grant No. 23qnpy58. This research is supported by the Associate Fellow Program from the APCTP.

\appendix

\section{Galaxy number-count dipole from super-horizon perturbation\label{app:A}}
In this appendix we give a rough proof that the super-horizon perturbation does not give sizable galaxy number-count dipole.  In the estimation of the galaxy number-count dipole, we adopt the results from previous studies \cite{PhysRevD.84.063505, DiDio:2013bqa, Nadolny:2021hti, Maartens:2017qoa}, which are also consistent with the findings of the authors in \cite{Domenech:2022mvt}. The expression for the dipole is given by
\begin{eqnarray}
&&\Delta({\bf n},z,m_*) = -\left( 2+\frac{\mathcal H^\prime}{\mathcal H^2} +\frac{2-5s}{r_S\mathcal H}- f_{\rm evo} \right) {\bf V_0\cdot n} \nonumber \\
&&+ b \delta + (f_{\rm evo}-3) \mathcal H V/ k - 3 \Phi \nonumber \\
&&+ (1+5s)\Phi+\Psi + \frac{1}{\mathcal H} \left[ \Phi^\prime + \partial_r({\bf V\cdot n}) \right] \nonumber \\
&&+ \left( \frac{\mathcal H^\prime}{\mathcal H^2} +\frac{2-5s}{r_S\mathcal H}+5s- f_{\rm evo} \right) \left( \Psi + {\bf V\cdot n} \right .\nonumber \\
&& \left .+ \int_0^{r_S} dr(\Phi^\prime+\Psi^\prime) \right)\nonumber \\
&&+ \frac{2-5s}{2 r_S} \int_0^{r_S} dr\left[2 - \frac{r_S-r}{r} \Delta_\Omega \right](\Phi+\Psi) ~.
\end{eqnarray}
In the equation above, ${\bf V_0\cdot n}$ term represents the velocity induced by super-horizon fluctuations. This term only contributes to the dipole component and is typically neglected in studies such as \cite{PhysRevD.84.063505, Nadolny:2021hti}. The quantities $f_{\rm evo}$ and $s$ are defined as follows:
\begin{eqnarray}
f_{\rm evo} &=& \frac{\partial \ln \left(a^3N(z, L>L_*)\right)}{\mathcal H \partial \tau_S}~, \\
s &=& -\frac{2}{5}\frac{\partial \ln N(z, L_S>L_*)} {\partial L_S} ~,
\end{eqnarray}
where $N(z, L>L_*)$ represents the number of galaxies with luminosity greater than $L_*$ at redshift $z$, and $L_S$ is the luminosity associated with the comoving scale $r_S$. In addition, $\delta$ is the matter density contrast in the co-moving frame, and $b$ is the linear bias parameter that connects the galaxy number-count fluctuations to the dark matter fluctuation. In this case, we assume $b=1$ for simplicity. $V, \Phi$ and $\Psi$ are the gauge invariant velocity and Bardeen potentials. 

The gauge-invariant velocity $V$, $\Phi$ and $\Psi$ and $\delta$ are related by the following expressions:
\begin{eqnarray}
\Psi&=&-\frac{3}{4} \left( \frac{k_{\rm eq}}{k}\right)^2\frac{1+y}{y^2} \delta~, \\
V&=& -\frac{3}{k} \mathcal H \frac{1+y}{4+3y}\left[ y \frac{d \delta }{dy}-\frac{1}{1+y} \delta \right]~,
\end{eqnarray}
where $y=a/a_{\rm eq}$ is a parameter related to the scale factor and $k_{\rm eq}$ represents the scale of equality between matter and radiation. It is worth noting that for super-horizon perturbations ($k \ll H_0$), the anisotropy of the stress can be ignored, leading to the relation $\Phi = \Psi$.

For adiabatic perturbation, assuming initial primordial power spectrum $\mathcal P(k)$ defined as
\begin{eqnarray}
\langle \Psi^{ini}_{\bf k}\Psi^{ini}_{\bf k^\prime} \rangle= 2\pi^2 \frac{\mathcal P(k)}{k^3} \delta({\bf k -k^\prime})~.
\end{eqnarray}
Then the power spectrum $C^N_l$ can be calculated by:
\begin{eqnarray}
C^N_l(z)= 4\pi \int\frac{dk}{k} \mathcal P(k) F^2_l~,
\end{eqnarray}
where 
\begin{eqnarray}
F_l &=& j_l (k r_S) \left[ b T_\delta + \left( \frac{\mathcal H^\prime}{\mathcal H^2} +\frac{2-5s}{r_S\mathcal H}+5s- f_{\rm evo}+1 \right) T_\Psi \right . \nonumber \\
&+& \left . (-2+5s) T_{\Phi} + \mathcal H^{-1} T^\prime_{\Phi} \right] \nonumber \\
&+& \left[ j^\prime_l(k r_S) \left( \frac{\mathcal H^\prime}{\mathcal H^2} +\frac{2-5s}{r_S\mathcal H}+5s- f_{\rm evo}\right) + j^{\prime\prime}_l(k r_S)\frac{k}{\mathcal H} \right . \nonumber \\
&+& \left . (f_{\rm evo}-3) j_l(k r_S) \frac{\mathcal H}{k} \right] T_V(r_S) \nonumber \\
&-& j^\prime_l(0) \left( 2+\frac{\mathcal H^\prime}{\mathcal H^2} +\frac{2-5s}{r_S\mathcal H}- f_{\rm evo}\right) T_V(0) \nonumber \\
&+& \frac{2-5s}{2r_S} \int_0^{r_S} \left[ (T_\Phi+T_\Psi) \left( l(l+1) \frac{r_S-r}{r} + 2 \right) \right]    \nonumber \\
&+& \int_0^{r_S}(T^\prime_\Phi+T^\prime_\Psi) \left(\frac{\mathcal H^\prime}{\mathcal H^2} +\frac{2-5s}{r_S\mathcal H} +5s- f_{\rm evo}\right) ~, 
\label{fleq}
\end{eqnarray}
where $j_l(x)$ is the spherical Bessel function, $T_\delta$, $T_\Psi$, $T_\Phi$, and $T_V$ are transfer functions, and $r_S$ represents the comoving distance of the galaxy sources. The equation above provides a way to compute the power spectrum $C^N_l$ at a given redshift $z$ by integrating over the primordial power spectrum $\mathcal P(k)$ and evaluating the function $F_l$ using the transfer functions $T_\delta$, $T_\Psi$, $T_\Phi$, and $T_V$.

The evaluation of the transfer functions needs numerical calculation. For super-horizon modes with $k < H_0$ and $y \gg 1$, the transfer functions can be analytically approximated as follows (from reference \cite{Hu:1994jd}):
\begin{eqnarray}
T_\Psi&=& \frac{9}{10}~, \nonumber \\
T_\Phi&=&T_\Psi~, \nonumber \\
T_V&=&\frac{1}{3} k \eta T_\Psi~, \nonumber \\
T_\delta&=&-\frac{4}{3} y \left( \frac{k}{k_{\rm eq}}\right)^2 T_\Psi~.
\end{eqnarray}
These expressions provide analytical approximations for the transfer functions $T_i$ ($i = \delta, V, \Phi, \Psi$) during the matter-dominated stage for super-horizon modes.

To check the result from the paper~\cite{Domenech:2022mvt}, we take $z=1$ (corresponding to $r_S=$ 3.4 Gpc) as an example, since the average redshift of quasars used in~\cite{Wright:2010qw} is around 1. We fix $k$ to satisfy $k r_{\rm dec} = 0.1$ and take $l=1$. We find that
\begin{eqnarray}
F_l(l=1)= - 10^{-5}-3\times 10^{-7} f_{\rm evo}- 10^{-5} s~. 
\end{eqnarray}
There must be large cancellations between different terms, as many terms are orders of magnitude larger than $F_l$. For example, the value of the $j^\prime_l(k r_S)$ term in the third line of Eq.~(\ref{fleq}) is, 
\begin{eqnarray}
(j^\prime_l(k r_S)~ {\rm term} )(l=1)= 0.02 -0.0076 f_{\rm evo}- 0.02 s ~. \nonumber \\
\end{eqnarray}
As demonstrated by the authors~\cite{Domenech:2022mvt}, for a single super-horizon mode, the cancellation occurs at the order of $k r_S$, and the main contribution is approximately at the order of $(k r_S)^3$.

In the case of a super-horizon isocurvature perturbation, as described in~\cite{Hu:1994jd}, we have the following relations,
\begin{eqnarray}
\Phi_{\bf k}&=&\Psi_{\bf k}= -\frac{1}{5}S_{\bf k}~, \nonumber \\
V_{\bf k}&=&-\frac{1}{15} k \eta S_{\bf k} ~, \nonumber \\
\delta_{\bf k}&=&\frac{4}{15} y \left( \frac{k}{k_{\rm eq}}\right)^2 S_{\bf k} ~.
\end{eqnarray}

To calculate $C^N_l$, we need to replace $\mathcal P$ with $\mathcal P_S$, and the transfer functions become,
\begin{eqnarray}
T_\Psi&=& -\frac{1}{5}~, \nonumber \\
T_\Phi&=&T_\Psi~, \nonumber \\
T_V&=&\frac{1}{3} k \eta T_\Psi \nonumber~, \\
T_\delta&=&-\frac{4}{3} y \left( \frac{k}{k_{\rm eq}}\right)^2 T_\Psi~.
\end{eqnarray}
Since the relation of $\Phi, V, \delta$ with $\Psi$ does not change, the cancellation in $F_l$ still exists. As shown in~\cite{Domenech:2022mvt}, such cancellation also holds for a single mode even if the dark energy-dominated stage is considered. For the continuum power spectrum, a full numerical calculation of the galaxy number-count dipole including the dark energy-dominated stage is needed, but it is beyond the scope of this paper and we leave it for future study. In the end, we provide the predicted galaxy number-count dipole for our benchmark point, assuming the universe is only matter-dominated. By considering only the super-horizon mode ($k < H_0$), assuming $f_{\rm evo}=1, s=1$ we find $C_1^N=6\times10^{-14}$, corresponding to a dipole around $6\times 10^{-8}$. Note that the sub-horizon mode also contributes to the dipole, but since $\mathcal P_S < \mathcal P_R$ (which is approximately $2.5\times 10^{-9}$) and it has been shown that the adiabatic perturbation $\mathcal P_R$ contributes only around $0.001$ to the galaxy number-count dipole at $z=1$~\cite{Nadolny:2021hti}, we believe the contribution of the galaxy number-count dipole from the sub-horizon isocurvature mode is negligible in our model.

\bibliography{bibly}

\end{document}